\documentclass[a4paper,11pt]{article}
\usepackage{amsmath}
\usepackage{amsthm}
\usepackage{amsfonts}
\usepackage{amssymb}
\usepackage{geometry}
\usepackage[parfill]{parskip} 
\usepackage{graphicx}
\usepackage{epstopdf}
\usepackage[colorlinks=true, pdfstartview=FitV, linkcolor=blue, citecolor=blue, urlcolor=blue]{hyperref}
\usepackage[matrix,frame,arrow]{xy}
\usepackage{booktabs}
\usepackage{tikz}
\usetikzlibrary{arrows,shapes,calc}

\input{Qcircuit}

\def\ket#1{|#1\rangle}
\def\cG{\mathcal{G}}

\DeclareMathOperator{\poly}{poly}

\newtheorem{theorem}{Theorem}
\newtheorem{corollary}{Corollary}

\newtheorem*{rep@theorem}{\rep@title}

\newcommand{\newreptheorem}[2]{%
\newenvironment{rep#1}[1]{%
 \def\rep@title{#2 \ref{##1}}%
 \begin{rep@theorem}}%
 {\end{rep@theorem}}}

\newreptheorem{theorem}{Theorem}

\newtheorem{defn}{Definition}
\newcommand{\secref}[1]{Section~\ref{sec:#1}}
\newcommand{\thmref}[1]{Theorem~\ref{thm:#1}}

\newcommand{\eqnref}[1]{Eqn.~(\ref{eqn:#1})}

\newcommand{\red}[1]{\textcolor{black} {#1}}

\begin{document}

\title{Efficient Distributed Quantum Computing}
\author{%
  Robert Beals$^1$, %
  Stephen Brierley$^2$\footnote{electronic address: steve.brierley@bristol.ac.uk}, %
  Oliver Gray$^2$\footnote{electronic address: oliver.gray@bristol.ac.uk}, %
  Aram W. Harrow$^3$, \and%
  Samuel Kutin$^1$, %
  Noah Linden$^4$, %
  Dan Shepherd$^{2,5}$ and %
  Mark Stather$^5$ \\ \\
$^1${\it Center for Communications Research, 805 Bunn Drive,}\\ {\it Princeton, NJ 08540, USA}\\
$^2${\it Heilbronn Institute for Mathematical Research, Dept. of Mathematics,} \\{\it University of Bristol, Bristol BS8 1TW, UK}\\
$^3${\it Dept. of Computer Science \& Engineering, University of Washington,} \\ {\it Seattle, WA 98195, USA}\\
$^4${\it Dept. of Mathematics, University of Bristol, Bristol BS8 1TW, UK}\\
$^5${\it CESG, Hubble Road, Cheltenham, GL51 0EX, UK}
}

\maketitle

\begin{abstract}
We provide algorithms for efficiently moving and addressing quantum memory in parallel.  These imply that the standard circuit model can be simulated with low overhead by the more realistic model of a distributed quantum computer.  As a result, the circuit model can be used by algorithm designers without worrying whether the underlying architecture supports the connectivity of the circuit. 
In addition, we apply our results to existing memory intensive quantum algorithms.  We present a parallel quantum search algorithm and improve the time-space trade-off for the Element Distinctness and Collision Finding problems.
\end{abstract}

\section{Introduction}  \label{sec:intro}

There is a significant gap between the usual theoretical formulation of quantum algorithms and the way that quantum computers are likely to be implemented.  
Descriptions of quantum algorithms are often given in the quantum circuit model in which, for an $N$-qubit circuit, there are up to $N/2$ simultaneous two-qubit gates in any one time-step. The model allows arbitrary interactions between qubits even though any implementation is likely to be mostly local in two or three dimensions, with a small number of long-range connections.  This is unlike classical computers, whose algorithms and implementations are currently dominated by von Neumann architectures where a small number of cores share access to a large random access memory (RAM).

On the other hand sometimes the circuit model is also less powerful than we would like. The concept of RAM or an analogue of the classical parallel RAM model would be useful for algorithm design. We introduce the quantum parallel RAM model which, in addition to any step of the circuit model, allows simultaneous queries to a shared quantum RAM. This allows us to design new quantum algorithms for parallel database searching, and the Element Distinctness and Collision Finding problems. 

We demonstrate that a single idea---\emph{reversible sorting networks}---can be used to efficiently relate the circuit and parallel RAM models of computation to one that is more physically realistic. We provide an efficient scheme for implementing a quantum circuit on a physical device where the possible pairwise qubit iterations are restricted. We then present a quantum circuit for implementing the quantum parallel RAM model, establishing that quantum computers can efficiently access memory in a parallel and unrestricted way. Combining these two results means that quantum memory does not need to be in a single place, but can be distributed amongst many processors.  

Our algorithms allow us to relate any quantum algorithm presented in terms of a quantum circuit to a distributed quantum computer in which each processor acts on a few well-functioning qubits connected to a small number of other memory sites (possibly via long-range interactions). At the most basic level a `memory site' could be a single qubit. Thus experiments such as NV centres in diamond, and trapped ions connected using optical cavities \cite{cirac+99, Monroe+12, Ritter+12}, or cavity QED for superconducting qubit networks \cite{blais+05, Wallraff+04,vanMeter+10}, could be used to efficiently implement quantum algorithms presented in the circuit model. 

Even in architectures strictly constrained to nearest-neighbour interactions in a constant number of dimensions, our results provide efficient simulations of the circuit model. For example, suppose we wish to implement a circuit with many concurrent operations on a 1D nearest-neighbour machine. One approach would be to use SWAP gates to bring each pair of qubits together so that the gate can be performed locally. For highly parallel circuits this would require as many as $O(N^2)$ time-steps since there are $N/2$ gates involving qubits that could be as far as $O(N)$ qubits apart. Using sorting networks (in this case the Insertion sort) Hirata et al. show how to perform all $N/2$ gates in only $O(N)$ time \cite{hirata}. This is a special case of our results for the 1D nearest neighbour architecture. Additionally, we prove that this is the best you can do; a 1D nearest neighbour machine cannot do better than emulate an arbitrary circuit with a $O(N)$ overhead.

Our work suggests to experimentalists how to begin building architectures for quantum computers. There is little need to build a quantum computer in which every qubit has the ability to interact with every other qubit. The addition of only a few long-range (flying) qubits allows the connectivity of a hypercube. There are efficient sorting networks over the hypercube (such as the Bitonic sorting network), so the overhead for emulating arbitrary quantum circuits on such a device is small.

Implicit in our work is the assumption of {\em full parallelism.}  In classical computing, this is generally unrealistic because communication is expensive and storage elements can be much cheaper than computation elements.  It may be that a passive fault-tolerant quantum memory will later be invented which will make storage costs of quantum memory substantially lower than the costs of performing computation.  However, it remains an open question whether such memories are physically possible in fewer than four spatial dimensions.  Instead, every known scheme for FTQC (fault-tolerant quantum computing) requires essentially full parallelism, even if the goal is simply to store quantum information without performing any computations.  Thus in this work, we will consider ``space'' and ``number of computing elements'' to be roughly synonymous.

Returning to the more theoretical aspects of our work, the results can be summarized as relating the following three models of quantum computation presented in Table \ref{Qmodels}: the well-known circuit model (Q circuit), quantum parallel RAM (Q PRAM), and a more realistic distributed quantum computing model (DQC) (also called a quantum multi-computer by van Meter et al.~\cite{vanMeter+06,vanMeter+06a}).  We prove that these three models are equivalent up to polylogarithmic depth overhead, as long as the DQC model is defined with respect to a sufficiently connected graph, such as the hypercube. 
 See Theorems~\ref{thm:UpperBound} and \ref{thm:LowerBound} and Corollary \ref{cor:bitonic} for explanations of how the DQC model can simulate the circuit model. Theorems \ref{thm:lower bound} and \ref{thm:main} give a precise description of how the circuit model can simulate a machine with quantum parallel RAM.

\begin{table}[ht]
\begin{tabular}{lp{0.4\textwidth}p{0.4\textwidth}}
\toprule
Model & In each time step & Limitations/Features \\
\midrule
Q circuit & Up to $N/2$ two-qubit gates are performed on arbitrary
disjoint pairs of qubits \vspace{1em}
 & No physical implementation since it requires all qubits to interact. Not clear how to perform simultaneous memory look-ups \vspace{1em} \\ 
Q PRAM & One step of circuit model {\em or} many simultaneous queries to
a shared RAM \vspace{1em} & 
RAM should be locally writeable (i.e. address $i$ controlled by processor
$i$) but globally accessible
\vspace{1em} \\ 
DQC & {$N$ small processors interact in a fixed
  low-degree graph (see Tab. \ref{tab:example graphs} for examples)}
 &{Can simulate the above two models. With degree $O(\log N)$ the overhead is $O(\log^2 N)$} \\
\bottomrule
\end{tabular}
\caption{Three models of quantum computation.}
\label{Qmodels}
\end{table}

Theorems 2 and 3 on simulating circuits apply after the quantum error
correcting layer so that the graph, $\cG$, in the DQC model consists of
logical qubits and gates (layer 4 in ref \cite{jones+12}). The emulation
process plays a role similar to the Solvay-Kiteav theorem \cite{kitaev97}
for approximating arbitrary single qubit gates given a universal gate set.
A general quantum algorithm in the circuit model may
involve arbitrary two-qubit interactions. We efficiently map these
gates to a sequence of logical gates restricted to $\cG$.

Our main result is \thmref{main} for efficiently accessing quantum memory in parallel.
The key primitive we need to implement is the following: Given
registers $\ket{i_1,\ldots,i_N}$, $\ket{x_1,\ldots,x_N}$ and
$\ket{y_1,\ldots,y_N}$ with $\ket{i_j, x_j, y_j}$ controlled by
processor $j$, we would like to replace the $\ket{y_j}$ register with
$\ket{y_j \oplus x_{i_j}}$.  In other words, processor $j$ has
specified an index $i_j$ and would like to query the $x_1,\ldots,x_N$
string at this position, although this string is distributed across
$N$ processors.
We achieve this using sorting networks, which are methods of sorting
$N$ objects via a series of pairwise reorderings of the form ``If
$X<Y$ then swap $X$ and $Y$.''  Such networks exist on the $N$-vertex
hypercube using $O(\log^2 N)$ depth~\cite{bitonic} and in many other
graphs (see Tab. \ref{tab:example graphs} or ref \cite{AKS}).  We can implement a sorting network reversibly by saving a bit for each
comparison to record whether a swap took place. We stress that no assumptions are made about the distribution of query indices, and that the algorithm works equally well if the queries are random or concentrated on a single node.

Our results apply when the processors are connected via a general graph with a topology-dependent overhead
factor.  The key consideration is the required depth of a sorting
network.  However, in order to reduce the number of qubits required to
save ``which path'' information, in some cases it is advantageous for
a weakly-connected architectures (such as a 2-d grid) to first
simulate a highly-connected architecture (such as a hypercube).  Doing
so would allow a $\sqrt{N}\times\sqrt{N}$ grid of qubits to simulate
the Q PRAM model with $O(\sqrt{N}\poly\log{N})$ depth overhead, which is
optimal up to the $\poly\log{N}$ factor.



\medskip \noindent {\bf Element Distinctness.}  An important
application of the quantum PRAM model and the efficient circuit for its implementation (\thmref{main}) is to quantum algorithms that
use large amounts of memory.   For example, the Element Distinctness
problem asks whether a function $f:[N]\rightarrow [N]$ is one-to-one or has
$x\neq y$ such that $f(x)=f(y)$, in which case output a colliding pair $(x,y)$. By guessing pairs of inputs this can be solved with a classical computer with
space $S$ and time $T$ constrained by $ST=O(N^2)$.  Observe that a naive application of Grover's algorithm \cite{Grover96}
fails to achieve any improvement in the number of queries: there
are $O(N^2)$ pairs to search, requiring $O(N)$ queries.
There is an elegant
quantum algorithm by Ambainis~\cite{ambainis07} that achieves the
optimal query complexity of $O(N^{2/3})$.  However, it also requires
memory $N^{2/3}\log N$; i.e. $S=\tilde{O}(N^{2/3})$, where the tilde notation implies that we have neglected factors polynomial in $\log N$.

When the function, $f$, is cheap to evaluate, but cannot be reverse-engineered, the best method of finding a collision is to treat $f$ as a black box. In this case, the same performance as Ambainis's algorithm can be achieved by a much simpler approach~\cite{grover+rudolph}.  We divide the $N^2$ search space into $N^2/S$ blocks and Grover search each one in time
$\tilde{O}(\sqrt{N^2/S})$ for a time-space tradeoff of $ST^2=\tilde{O}(N^2)$, which includes
 $S=T=\tilde{O}(N^{2/3})$ as a special case. Grover and Rudolph pose beating this trade-off as a challenge.

Our main result (\thmref{main}) allows us to meet this challenge. We
choose a random set of $S$ inputs $x_1,\ldots,x_S$ to query, and
store these along with $f(x_1),\ldots,f(x_s)$ in the
distributed memory using a data structure that permits fast lookups.
Next, we use the $S$ processors to check the 
remaining $N-S$ inputs for a collision with one of the
$x_1,\ldots,x_S$.  Using Grover's algorithm $S$ times in parallel (\thmref{multiGrover}), this requires time $\tilde{O}(\sqrt{N/S})$.
Note that each iteration of Grover requires the use of our algorithm for making
simultaneous queries to a distributed memory.

This will succeed (with constant probability) in finding a collision
if one of the colliding elements was in $\{x_1,\ldots,x_S\}$.  Since
this event happens with probability $\Omega(S/N)$ (as long as a collision exists at all), we can wrap this
entire routine in amplitude amplification to boost the probability to
$2/3$ by repeating $O(\sqrt{N/S})$ times.   The total time is
$T=\tilde{O}(N/S)$, and so we have achieved a time-space tradeoff of 
\[ ST = \tilde{O}(N) \, .\]
We stress that due to the efficient circuit for parallel memory look-ups (\thmref{main}), the trade-off applies in the circuit model where space is the number of qubits (or width) and time is the circuit depth.

\medskip \noindent {\bf Organisation of the paper.} The remainder of the paper is organised as follows. In Section \ref{sec:processors}, we provide an algorithm for efficiently moving quantum data (\thmref{move}) according to a \emph{classical} permutation. Distinct indices are permuted, replacing $\ket{x_1,\ldots,x_N}$ with $\ket{x_{j_1},\ldots,x_{j_N}}$. Armed with \thmref{move}, we relate the circuit model to the distributed quantum computing model in \secref{DQC}. 
We prove that given (by an experimentalist) \emph{any} layout of qubits grouped into memory sites with connections between fixed pairs of sites, a device with this layout can implement algorithms presented in the circuit model.  Of course there is a price to pay: the overhead depends on the topology of the processors, but our algorithm is close to optimal. 

In Section \ref{sec:defs}, we generalise the data-moving algorithm so that the destinations $j_1, \ldots, j_N$ are \emph{quantum} data and no longer restricted to form a permutation. This results in an algorithm (circuit) that can look up multiple memory entries in parallel. In addition, we prove that this memory look-up algorithm (measured in terms of circuit complexity) is scarcely more expensive than \emph{any} circuit capable of accurately accessing even a single entry. Thus, the quantum parallel RAM model can be efficiently simulated using a distributed quantum computer.  We consider, in \secref{algs}, various quantum search problems in the quantum parallel RAM model. The application of our efficient algorithm for parallel memory access results in several improved quantum algorithms: The Multi-Grover algorithm for parallel search of an unstructured database; an improved algorithm for the Element Distinctness problem and an improved algorithm for the Collision Finding problem.

\section{Moving Quantum Data}  \label{sec:processors}

The quantum state stored in a qubit can be moved across a graph, $\cG$, using a chain of SWAP gates. Alternatively, we could apply a teleportation scheme making use of existing entanglement and classical control \cite{cleve+97}.  However, if in a single phase of computation, we wish to perform a permutation of many or even all of the qubits, this routing problem is non-trivial. 

We consider the problem of moving $N$ pieces of quantum data, initially stored at the $N$ nodes of $\cG$. In this section we provide an algorithm, $V_N$, that performs an arbitrary permutation of qubits. Our only requirement being that the host graph, $\cG$, supports a sorting network. We make two simplifying assumptions: that the destinations are known at the outset (as classical information) and that they are all different. In \secref{defs}, we relax these assumptions and generalise our algorithm.

\begin{defn}  \label{def:dm}
Given the quantum register $\ket{x_1,x_2,\ldots,x_{N}}$ with $\ket{x_j}$ located at node $j$ of the graph $\cG$, an algorithm for \emph{data moving} is a unitary map implementing
\begin{eqnarray}
   V_N ~:~ \ket{x_1,x_2,\ldots,x_{N}}   \mapsto  \ket{x_{j_1},x_{j_2},\ldots,x_{j_N}}, \nonumber
\end{eqnarray}
for some permutation $(1,\ldots,N)\mapsto (j_1,\ldots,j_N)$. The $N$ data registers $x_1,\ldots,x_{N}$ each consist of $d$ qubits.
\end{defn} 

We now present an efficient algorithm for implementing $V_N$. The result applies to \emph{any} connected graph and we give some interesting examples and their respective properties in Table \ref{tab:example graphs}. 
\begin{theorem} \label{thm:move}
Let $\cG$ be a connected graph with $N$ nodes each with $d$-qubit data registers and $Q$-qubit ancilla space. Let $D_\cG$ be the depth of a reversible sorting network over $\cG$ with $N$ inputs. Then there exists an algorithm for $V_N$, restricted to $\cG$, with depth $O(D_\cG)$.
\end{theorem}

Before proving \thmref{move}, we first describe the main subroutine for the data-moving algorithm and subsequent algorithm for parallel memory access: a classical reversible sorting network. Both the data-moving and parallel memory access algorithms are in fact \emph{classical reversible circuits}.  Indeed, the tools we use were developed originally for classical parallel computing \cite{Valiant88,Valiant90,Leighton92}.  We were unable to find our Theorems~\ref{thm:move} or~\ref{thm:main} in the literature (although \cite{Valiant88} is somewhat similar), and we suspect that may be because quantum computing presents architectural issues that are different to those encountered in the field of traditional classical parallel computing.

\begin{table}[t]
\center
\begin{tabular}{p{0.3\textwidth}c p{0.3\textwidth}c}
\toprule
Graph & Valency & Sorting network & $D_\cG$\\
\midrule
1D Nearest Neighbour & $2$ \vspace{0.5em} & Bubble/Insertion sort \vspace{0.5em}  & $O(N)$  \vspace{0.5em} \\ 
2D Nearest Neighbour & $4$ \vspace{0.5em} & Bubble/Insertion sort \vspace{0.5em}  & $O(\sqrt{N})$  \vspace{0.5em} \\ 
Hypercube & $O(\log N)$& Bitonic sort \cite{bitonic} & $O(\log^2 N)$   \\
 &  \vspace{0.5em} & AKS sort \cite{AKS} \vspace{0.5em}  & $O(\log N)^*$  \vspace{0.5em} \\
Complete graph & $N$ & Any sorting algorithm & $O(\log N)$  \\
\bottomrule
\end{tabular}
\label{tab:example graphs}
\caption{Comparision of various $N$ node graphs giving the graph valency and examples of sorting networks over $\cG$. The depth of the sorting network, $D_\cG$, translates into the cost of running $V_N$ via Theorem \ref{thm:move}. ($^*$ includes a large constant $c\approx 6,1000$.)}
\end{table}

\subsection{Sorting over a graph} \label{sec:sort}

A \emph{sorting network} is a network on $T$ \emph{wires} in which each wire represents one of $T$ elements and where the only gates are \emph{binary comparators}.  A binary comparator takes two elements on its input, and it outputs the same two elements but in the correct order, according to some specified comparison routine.  Crucially, a sorting network is independent of the inputs, the same fixed set of gates correctly sort all inputs. 

For any $T>0$ there must be a lowest-depth sorting network for sorting $T$ elements deterministically.  More generally, one can say that for any connected graph $\cG$ of $T$ vertices, there must exist a lowest-depth sorting network for sorting $T$ elements, all of whose comparators lie along edges of $\cG$.
For example, when $T=4$ and $\cG$ is a 4-cycle, then the bitonic sorting network fits the graph, and, having six comparators in depth 3, is optimal (cf. \secref{bitonic}).  But for $T=4$ and $\cG$ a graph of three edges in a line, the bitonic sort is not possible, and a sort involving six comparators in depth 4 turns out to be optimal for this graph.
Examples of sorting networks for popular architectures are given in Table \ref{tab:example graphs}.

To make a sorting network reversible, we add an additional \emph{sorting-ancilla} bit to each comparator that is initially set to $\ket0$.  This sorting-ancilla gets flipped to a $\ket1$ if and only if the comparator exchanges the order of the two elements input.  In more detail, the comparator \emph{first} compares the two elements, storing the resulting bit in the sorting-ancilla; \emph{then} conditioned on the sorting-ancilla being set, it swaps over the two elements.  These two steps are each clearly reversible in their own right.  The sorting-ancillas are then all retained, to enable `unsorting' later. 


A reversible comparator for full lexicographic sort on $b$-bit objects was constructed by Thapliyal et al. \cite{Thapliyal+10}. Their algorithm is based on the binary tree and is  efficient for our purposes having width $O(b)$ and depth $O(\log b)$.

Write $s(T)$ for the total number of comparators appearing in a given sorting network for $T$ elements.  A sorting operation is written as follows:
\begin{eqnarray}  \label{eqn:sort}
  S_T ~:~ \ket0 \otimes \bigotimes_{k=1}^{T} \ket{x_k}
  &\mapsto&  \ket\sigma \otimes \bigotimes_{k=1}^{T} \ket{x_{\sigma(k)}}.  
\end{eqnarray}
Here $\sigma \in \text{Sym}(T)$ denotes a permutation that puts the $T$ elements in order, and the first register (storing $\sigma$) actually consists of the $s(T)$ sorting-ancilla bits that get fed into the comparators.  
It is clear from this observation that $s(T) \ge \log_2 (T!) = \Omega( T \log T)$ for any valid sorting network, since potentially any element of $\text{Sym}(T)$ might be a uniquely correct one.  If a sorting network has depth $D$, then it uses a total of $O( D \cdot T )$ comparators, and so depth must be $D=\Omega( \log T )$ for any valid sorting network.

Because the sorting subroutine is reversible, it makes sense to run it backwards.  When that happens, the comparators are encountered in reverse order, and each comparator swaps the order of its inputs according to whether its sorting-ancilla bit is set.  That sorting-ancilla bit is then reversibly cleared (regardless of its value) by `uncomputing' the comparison between the two elements.
\subsection{The quantum data-moving algorithm}  \label{sec:defV}

\def\cheese#1{\widehat{#1}}

We now describe the quantum data-moving circuit, $V_N$.
It is convenient to break $V_N$ into three basic parts: formatting, sorting, and applying the permutation.  The formatting and sorting both need to be reversed after the permutation has been applied.  
This can all be annotated as follows, reading right to left:
\begin{eqnarray}  \label{eqn:defVN}
  V_N  &=&  \cheese{F} \circ S_{2N}^{-1} \circ P \circ S_{2N} \circ F.
\end{eqnarray}
We must be careful to ensure that all operations can be performed with the graph locality restriction and to count the depth and additional ancilla space required by each of these subroutines (see~\secref{count}).
The format and permutation will both be entirely local operations, so their circuits are already admissible to the processor model. The gates of the sorting subroutine all belong to comparators of the sorting network over the graph, $\cG$. 

\paragraph{1) Initial Formatting}

The subroutine $F$ can be achieved using Pauli $X$ gates and SWAP gates with no additional ancillas and in depth 1.  It is not really a `computation' at all, rather a rearrangement of the input data into a format amenable for describing the sorting that will follow.

Let there be $2N$ ancilla registers, which we call \emph{packets}. Each packet contains an \emph{address} ($\lceil \log_2 N \rceil$ bits), a \emph{flag} (1 bit), and a \emph{data} ($d$ bits) register.  The initial format moves the input data out of the input registers and into the packets, ready for sorting:
\begin{eqnarray}\label{eq:format}
  \lefteqn{ F ~:~ \ket{x_1,x_2,\ldots,x_{N}} \bigotimes_{i=1}^{N} \ket{(0,0,0)}_{2i}~\ket{(0,0,0)}_{2i+1} }  \phantom{XXXX} \\
  &\mapsto&  \ket{0,0,\ldots,0} \bigotimes_{i=1}^{N} \ket{(i,a,x_i)}_{2i}~\ket{(j_i,q,0)}_{2i+1},\nonumber   
\end{eqnarray}
where $j_r \neq j_s$ for all $r\neq s$.  The special symbols $q$ and $a$ denote `question' and `answer', respectively.

One can think of the map $F$ in the following terms. Each processor, $i$, submits the `answer' packet $(i,a,x_i)$, and if the data it would like to obtain is at index ${j_i},$ it also submits the `question' packet $(j_{i},q,0)$. The flag is used in the sort and unsort steps with the convention that $q<a$.

\paragraph{2) Sorting}

Sorting was described in \secref{sort}.  In this context, we want to sort the $2N$ packets, using a lexicographical ordering that reads only the address then the flag of each packet. In accordance with \eqnref{sort}, the subroutine $S_{2N}$ must employ a register of $s(2N)$ sorting-ancilla bits, mapping as 
\begin{eqnarray}  \label{eqn:sort2}
  \lefteqn{ S_{2N} ~:~ \ket0 \otimes \bigotimes_{i=1}^{N} \ket{(i,a,x_i)}_{2i}~\ket{(j_i,q,0)}_{2i+1} } \phantom{XXXX} \\
  &\mapsto&  \ket\sigma \otimes \bigotimes_{i=1}^{N} \ket{(i,q,0)}_{2i}~\ket{(i,a,x_i)}_{2i+1}, \nonumber  
\end{eqnarray}
where $\sigma \in \text{Sym}(2N)$ is the permutation implied by $2i = \sigma(2i+1)$ and $2i+1 = \sigma(2j_i)$, for $i$ from $1$ to $N$.

The total depth of the circuit for unitary $S_{2N}$ is equal to the depth of the sorting network multiplied by the depth of a single comparator.

\paragraph{3) Locally SWAP}

After the sort, we are left with a sequence of packets of the form
\begin{eqnarray}
\ldots(i-1,q,0)~(i-1,a,x_{i-1})~(i,q,0)~(i,a,x_{i})~(i+1,q,0)~(i+1,a,x_{i+1})\ldots 
\end{eqnarray}
where the packets are sorted in lexicographical order according to the address and the flag. This ordering means that each answer is immediately to the right of the corresponding question and, furthermore, both lie on the same node.  Without the need for any auxiliary bits, and in depth 1, SWAP gates can achieve the map
\begin{eqnarray}
  P ~:~ \bigotimes_{i=1}^{N} \ket{(i,q,0)}_{2i}~\ket{(i,a,x_i)}_{2i+1}
  &\mapsto&  \bigotimes_{i=1}^{N} \ket{(i,q,\red{x_i})}_{2i}~\ket{(i,a,\red{0})}_{2i+1}.
\end{eqnarray}
An important property of $P$ is that it does not change the address or the flag used in the sort.

\paragraph{4) Unsorting}

The sorting network can be run in reverse to return the packets to their original positions. To achieve the unsort, $S^{-1}_{2N}$ acts on the sorting-ancilla register and the packets, mapping
\begin{eqnarray}  \label{eqn:unsort2}
  \lefteqn{ S_{2N}^{-1} ~:~ \ket\sigma \otimes \bigotimes_{i=1}^{N} \ket{(i,q,x_i)}_{2i}~\ket{(i,a,0)}_{2i+1} } \phantom{XXXX} \\
  &\mapsto& \ket0 \otimes \bigotimes_{i=1}^{N} \ket{(i,a,0)}_{2i}~\ket{(j_i,q,x_{j_i})}_{2i+1}.  \nonumber
\end{eqnarray}
Since $S^{-1}_{2N}$ is the same as the sorting network, but with the order of the gates reversed, the cost for the unsort step is the same as for step 2.

\paragraph{5) Final formatting}
The final step in the algorithm is to write the data back to the original registers and clear the ancilla space. As with the initial formatting, $F$, the subroutine $\cheese{F}$ has depth 1.  Acting on the same registers as for its counterpart $F$, it works as follows:
\begin{eqnarray}
  \lefteqn{ \cheese{F} ~:~ \ket{0,0,\ldots,0} \bigotimes_{i=1}^{N} \ket{(i,a,0)}_{2i}~\ket{(j_i,q,x_{j_i})}_{2i+1} }  \phantom{XXXX}  \\
  &\mapsto&  \ket{x_{j_1},x_{j_2},\ldots,x_{j_{N}}} \bigotimes_{i=1}^{N} \ket{(0,0,0)}_{2i}~\ket{(0,0,0)}_{2i+1}.  \nonumber
\end{eqnarray}

\subsection{Low valency graphs}
We require $O(D_\cG)$ qubits at each node to record the which-way information of the two sorting steps in $V_N$. Hence for very low valency graphs where the sorting depth is large, there is insufficient local space to run the algorithm. For example, in the 1D nearest neighbour graph we would need each node to have $O(N)$ qubits. However in any physically relevant architecture, the node size is likely to be much smaller than $O(N)$.  We now explain how to fix this problem. 

The sorting step in the $V_N$ algorithm involves only the classical information about the permutation known at the start of the algorithm; it is independent of the data register.  We can classically compute all of the moves required by the sorting network before running the quantum algorithm. Hence we can replace the reversible comparators (with their additional storage qubit) by a SWAP gate or no gate as required in that step of the sort. Since the network remains the same, the depth of the algorithm is $O(D_\cG)$, as before, but with $O(d)$ qubits per node. This is the approach taken by Hirata et al. who consider the case of the 1D nearest neighbour graph with one qubit per site \cite{hirata}.

\section{Efficient Distributed Quantum computing}\label{sec:DQC}
The circuit model allows any pair of qubits to be connected by gates and so allows arbitrarily long-range interactions. This model is very far from any likely implementation of a quantum computer. We imagine that a small number of long-range interactions could be possible, but most gates will need to be local. Due to the presumed requirements of fault-tolerance \cite{ABIN}, we expect implementations will allow the concurrent execution of one- and two-qubit gates with fast classical control

It is well known that if the qubits are laid out in a line and each could only connect with its nearest neighbours (one either side), then the resulting model of computation would still be universal because it could emulate any circuit of the more general kind.  The emulation proceeds in a straightforward fashion using SWAP gates to bring qubits next to each other so that nearest-neighbour gates can be applied.  The price to pay in this emulation is (in general) an overhead factor of $O(W)$ for each gate, where $W$ counts the total width (number of qubits) of the circuit being emulated. For highly parallel algorithms there are O(W) gates per timestep so the cost of this emulation scheme is $O(W^2)$. Hirata et al. \cite{hirata} use a sorting algorithm to emulate all $O(W)$ gates simultaneously, resulting in an improved emulation scheme for the 1D nearest neighbour architecture with cost $O(W)$.

More generally, we could envisage having memory larger than a single qubit at each `site', with connectivity more generous than simply being connected each to two neighbours in a line.  Then the \emph{overhead depth factor} (or \emph{emulation factor}) of embedding circuit operations into the underlying hardware could be reduced to something smaller than $O(W)$.  
In this section we demonstrate that reversible sorting networks provide an elegant way of embedding an arbitrary circuit into physical hardware. Up to logarithmic factors, the proposed solution is asymptotically optimal. Our scheme is very general and includes the case of each site being a single qubit.

Consider a quantum circuit of width $W$, and let there be $N$ quantum processors, each with its own local memory of $Q$ qubits.  Suppose that $Q \cdot N \ge W$ and that the processors are interconnected as the $N$ nodes of a graph $\cG$.  We say that a circuit \emph{respects graph locality} if every two-qubit gate in the circuit has those two qubits lying \emph{either} in the same processor's local memory \emph{or} in a neighbouring memory with respect to $\cG$.  
In addition, we require that each gate is assigned to a processor that holds at least one of its qubits, and each processor can only perform one gate per time-step.
Together, these restrictions on the ordinary circuit model define the distributed quantum computing model formally.

We examine the best overhead depth factor for arbitrary circuit emulation in the worst case. Starting with a circuit of $W$ qubits, we wish to emulate it using a circuit that respects graph locality.  We want the overhead depth factor of this emulation to be as small as possible. That is, we wish to minimise the function
\begin{eqnarray}
  F(\cG,Q,W)  &:=&  \max_C ~\frac{Depth(C')}{Depth(C)} 
\end{eqnarray}
where $C'$ is a circuit for emulating $C$ subject to the constraints imposed by the host graph $\cG$ and the number of qubits at each site, $Q$.  Maximisation is over all circuits $C$ of width $W$ so $F(\cG,Q,W)$ is the worst case cost of emulating arbitrary circuits.
Normally we are concerned with the case $W=|\cG|=N$, so that an emulation has one processor per qubit being emulated, and where $Q$ is large enough to hold ancilla for basic computations.

In \secref{processors} we used sorting networks to efficiently move quantum data. We now show that scheduling tasks on a distributed quantum computer can be regarded as (more or less) equivalent to the problem of designing sorting networks commensurate with those topologies.  We put these ideas on a firmer footing in the following Theorem. 

\begin{theorem}  \label{thm:UpperBound}
Let $\cG$ be any connected graph on $N$ nodes, and consider the distributed quantum computing model with graph $\cG$ and $Q=O(\log N)$ qubits per processor.  Let $D_{\cG}$ denote the depth of the best algorithm for sorting $2N$ arbitrary bit strings of length $\lceil \log_2 N \rceil + 2$ over $\cG$.
Then there exists an algorithm for emulating arbitrary circuits with depth overhead given by 
\begin{eqnarray}  \label{eqn:upper}
  F( \cG, Q, N ) =  O( D_{\cG})
\end{eqnarray}
\end{theorem}

\emph{Proof~:~}
The proof of the Theorem uses our algorithm for $V_N$ with data registers of size $d=1$  (\emph{cf}.~\S\S\ref{sec:defV}). The circuit for $V_N$ can be directly embedded into the distributed quantum computer. Hence we show how a circuit of width $N$ can be emulated in terms of gates that are local (with respect to $\cG$) and renditions of $V_N$ only.

Each processor is large enough to hold two packets and some ancillas, since $Q = O(  \log N )$. Given a general circuit of width $N$, in each timeslice there will be at most $N$ gates. The gates are `assigned' one per processor when the timeslice is emulated.  When a processor comes to emulate a gate assigned to it, it will need access to the one or two qubits of that gate.  The emulation of a timeslice therefore requires two calls to the subroutine $V_N$: without loss of generality, we can assume that the first qubit of a gate already resides at the processor to which that gate has been assigned; the first call to $V_N$ brings the second qubit of each gate to its processor;  the processor implements the gate locally on the two qubits; the second call to $V_N$ restores the second qubit of each gate to its original home.  Null ancilla qubits can be included within each $V_N$ operation in order to make it a permutation.

Every timeslice of the circuit being emulated now additionally requires two calls to $V_N$, plus appropriate $O(\log N)$-sized circuitry (per processor) to write and erase the indices $j_i$ used within $V_N$.  Overall this costs an overhead depth factor of $O( D_{\cG} )$. 

This completes the proof of \thmref{UpperBound}.  
\hfill  \qed  \bigskip

In \thmref{UpperBound}, we provided an algorithm for embedding a circuit into a physical host graph. Our aim was to minimise the depth overhead factor $F(\cG, Q, N )$ of emulating the worst case circuit. In the following theorem, we prove that this algorithm is optimal up to logarithmic factors.
\begin{theorem}  \label{thm:LowerBound}
For a distributed quantum computer with host graph $\cG$ consisting of $N$ nodes each with $Q$ qubits, the cost of emulating arbitrary circuits of width $O(N\log N)$ is bounded by
\begin{eqnarray}  \label{eqn:lower}
 F_{min}(\cG, Q, O(N \log N) )   
 &=& \Omega\left(\frac{D_{\cG}}{\log N\log\log N}\right)  \, . 
\end{eqnarray}
\end{theorem}
\emph{Proof~:~}  
The AKS sorting network sorts $T$ elements in $O(\log T)$ comparator depth \cite{AKS, paterson90}. Suppose we use this network to sort $2N$ packets of bit-length $\lceil \log_2 N \rceil + 2$.  Let $C$ be the (unconstrained) circuit for achieving this, so the width of $C$ is $W=O(N \log N)$ and its depth is $O(\log N \log\log N)$ (\emph{cf.} \secref{count}).  The cost of emulating $C$ is bounded by $D_{\cG}$ since the emulation is a sorting algorithm on the $(\cG,Q)$-distributed computer and we defined $D_{\cG}$ to be the depth of the \emph{best} sorting algorithm. Hence the cost of any emulation scheme is lower bounded by $\Omega(D_{\cG}/\log N \log\log N)$, as required.
\hfill  \qed  

A lower bound on the cost of implementing an addition circuit on a $k$D nearest neighbour graph was found by Choi and Van Meter \cite{choi+08}. Their bound translates into an emulation cost of 
$$F_{min}(\cG, Q, N )=\Omega(N^{1/k})=\Omega(D_\cG)$$
in the special case of emulating a circuit for addition on a $k$D nearest neighbour graph.

\subsection{The Hypercube architecture}\label{sec:bitonic}
Theorem \ref{thm:UpperBound} provides an efficient algorithm for emulating circuits on \emph{any} graph. We briefly consider a particularly nice architecture, the hypercube, demonstrating that the addition of a few flying qubits greatly improves the efficiency of a distributed quantum computer. 

All of the comparators in any given sorting network are identical, and when any $T$ elements are input to the sorting network, its overall effect should be to output them in totally sorted order, with certainty.  
The network can of course be designed completely independently of the comparator, since the details of what makes one element `greater', `less', or `equal' to another is irrelevant from the perspective of which binary comparisons are needed to guarantee sorting. Thus for any value of $T$, one can ask for the lowest depth sorting network for sorting $T$ elements.  

In references \cite{AKS, paterson90}, it is shown that $T$ elements can be sorted in $O(\log T)$ depth of comparators, and that a uniform family of sorting networks achieves this. Unfortunately the constant for Paterson's simplified version of the AKS sorting network \cite{paterson90} is around $6100$ and so not practical for any realistic sizes of $T$. However, the \emph{bitonic sorting network} \cite{bitonic} sorts $T=2^t$ elements in depth $1+2+\hdots+t=t(t+1)/2=O( \log^2 T )$.

The following corollary of \thmref{LowerBound} demonstrates a nice balance between the cost of embedding arbitrary circuits into the graph, the size of the nodes, $Q$, and the degree, $val(\cG)$. 

\begin{corollary}  \label{cor:bitonic}
A distributed quantum computer with $N$ nodes of $Q = O(\log N)$ qubits interconnected using a hypercube architecture has the following properties: the number of connections per node is $val(\cG) = O(\log  N)$, and yet the overhead depth factor is $F(\cG,Q,N)=O(\log^2 N \log\log N)$ for emulating quantum circuits of width $N$.
\end{corollary}

Ignoring constants, the AKS graph would be a better graph to use having overhead depth factor $F(\cG,Q,N)=O(\log N \log\log N)$. Hence asymptotically, the cost of emulating the circuit model on a distributed quantum computer with low degree is $O(\log N)$. Alternatively, there are sorting algorithms with an $O(\log N)$ complexity but which are probabilistic and so solve the routing problem on almost all circuits. However, such a probabilistic approach will not work if we require \emph{all} permutations to be correctly routed. This is the case in the next section when we consider parallel memory access.

\section{The cost of accessing quantum memory}  \label{sec:defs}

In \secref{processors} we considered moving quantum data according to a permutation known at compile time. By using the permutation needed to embed a circuit into a host graph, this approach leads to an efficient algorithm for distributed quantum computing in \secref{DQC}. The algorithm for data moving can be easily modified to perform a permutation that is stored as \emph{quantum} data, $\ket{j_1, \ldots , j_N}$. We simply amend the formatting steps of the algorithm, using SWAP rather than $X$ gates
\begin{eqnarray}
  \lefteqn{ F ~:~ \ket{j_1, \ldots , j_N}\ket{x_1,x_2,\ldots,x_{N}} \bigotimes_{i=1}^{N} \ket{(0,0,0)}_{2i}~\ket{(0,0,0)}_{2i+1} }  \phantom{XXXX} \\
  &\mapsto&  \ket{0,0,\ldots,0} \bigotimes_{i=1}^{N} \ket{(i,a,x_i)}_{2i}~\ket{(j_i,q,0)}_{2i+1} \, . \nonumber   
\end{eqnarray}
Such an algorithm would allow us to move data in a superposition over permutations. In this section, we generalise the idea further to the case where the indices do not necessarily form a permutation but could be equal. This new algorithm allows efficient parallel access to quantum memory and leads to the new quantum algorithms presented in \secref{algs} .

\subsection{The cost of a single look-up}  \label{sec:2.1}
 
Often when quantum algorithms are quoted in the \emph{query model}, the concept of an \emph{oracle} is used to abstract away those logical unitaries that are intended to make `random' access to (quantum) memory.  We start by examining the idea of accessing memory in more detail, without using oracles.

\begin{defn}\label{defn:database}
A logical unitary for accessing a single piece of data is a map $U_{(1,N)}$ that implements 
\begin{eqnarray}
  U_{(1,N)} ~:~ \ket{j}\ket{y}\ket{x_1,x_2,\ldots,x_{N}}  &\mapsto&
    \ket{j}\ket{\red{y \oplus x_j}}\ket{x_1,x_2,\ldots,x_{N}},
\end{eqnarray}
where $\oplus$ denotes bitwise addition.
\end{defn}

Here we have depicted $2+N$ registers.  
The first register (\emph{index register}) holds an index large enough to `point' to any one of $N$ \emph{data registers}; its associated Hilbert space must be of dimension at least $N$.  
The second register is called the \emph{target register} and holds the same kind of data as a data register.
The other $N$ registers are data registers and could in principle be of any (equal) size.  

We can derive a simple lower bound for the cost of accessing a single piece of memory based on two simple constraints on any circuit for $U_{(1,N)}$. The memory must hold the entire database and so within the circuit, there must be a causal chain from every data register to the target register. This observation has been made before \cite{bernstein09}, but we give it here to motivate the cost of our parallel look-up algorithm that follows. 

\begin{theorem}  \label{thm:lower bound}
In the circuit model, if a circuit implements $U_{(1,N)}$ on $N$ data registers each consisting of $d$ qubits, then its width is $\Omega( N d )$ and its depth is $\Omega(\log N)$.
\end{theorem}

\emph{Proof~:~}
The width of the circuit must be $\Omega( N d )$ because this is the logical width of the unitary (the number of input/output qubits).  Since any data register could affect the target register, there must be a causal chain of gates from any data register to the target register.  Each permitted gate touches $O(1)$ registers, so the depth of the longest chain must be $\Omega( \log N  )$.
\hfill  \qed  \bigskip

There is a sense in which the gates in a typical circuit for $U_{(1,N)}$ can be said to be ``not working very hard'' (although this idea is hard to quantify precisely), and this inefficiency points to the need for a parallel algorithm.

%
%

\subsection{The cost of parallel look-ups}  \label{sec:2.2}

\begin{defn}  \label{def:pd}
A logical unitary for accessing N pieces of data is a map $U_{(N,N)}$ that implements
\begin{eqnarray}\label{eqn:pd}
  \lefteqn{ U_{(N,N)} ~:~ \ket{j_1,j_2,\ldots,j_{N}}\ket{y_1,y_2,\ldots,y_{N}}\ket{x_1,x_2,\ldots,x_{N}} } \phantom{XXXX}  \\
   &\mapsto&  \ket{j_1,j_2,\ldots,j_{N}}\ket{\red{y_1 \oplus x_{j_1},y_2 \oplus x_{j_2},\ldots,y_{N} \oplus x_{j_{N}}}}\ket{x_1,x_2,\ldots,x_{N}}. \nonumber
\end{eqnarray}
\end{defn}
Here we have depicted $N$ index registers, $N$ target registers, and $N$ data registers, comprising a total of $3N$ input registers.  As before, the index registers are each made up of $\lceil \log_2 N \rceil$ qubits, while the target registers and data registers are each made up of $d$ qubits.

We define the quantum parallel RAM (Q PRAM) model as the circuit model plus the ability to access quantum memory in a parallel and unrestricted way. In any one time-step we can apply up to $N/2$ two-qubit gates on arbitrary disjoint pairs of qubits in the first two registers, or we can apply $U_{N,N}$.  Previous version of quantum RAM such as~\cite{Giovannetti+08,Giovannetti+08b} do not consider parallel access.

In the following theorem we prove that the circuit model can emulate the Q PRAM model with only a logarithmic cost in the width and depth. The input value, $y_i$,  to $U_{(N,N)}$ can be interpreted as the current state of processor $i$ and the output value $y_i \oplus x_{j_i}$ as the result of a memory request by this processor to processor $j_i$.

\begin{theorem}  \label{thm:main}
There is a uniform family of quantum circuits implementing $U_{(N,N)}$ defined in Eq. (\ref{eqn:pd}). This circuit family has width $O(N (\log N + d) )$ and depth $O( \log N \log(d \log N) )$, where $d$ denotes the size of the data registers.
\end{theorem}

\emph{Proof~:~}The algorithm for $U_{(N,N)}$ is presented in \secref{descU}. The calculation of the width and depth can be found in Appendix A.
\hfill  \qed  \medskip


Theorems~\ref{thm:lower bound} and~\ref{thm:main} also tell us that in the circuit model, for parallel memory lookups, we can achieve a factor $N$ more `effect' for only a small additional `effort'.  This will be seen to have radical effects on certain `memory-intensive' algorithms in~\secref{algs}.

\subsection{The parallel look-up algorithm}  \label{sec:descU}

We now present the algorithm for accessing memory in parallel, implementing the unitary $U_{(N,N)}$ defined in \eqnref{pd}. 
As with $V_N$, a circuit for $U_{(N,N)}$ depends only on the parameters $N$ and $d$.  We will generalize the algorithm given for $V_N$ to show how $U_{(N,N)}$ may be efficiently implemented. 

The overall structure for $U_{(N,N)}$ is the same as for $V_N$. We first construct packets that include the original data and a flag to be used in the sorting step. After a sorting network is applied, we transform the data. Instead of the permutation used in the data moving algorithm, the transformation to use is composed of \emph{cascading}, $B$, \emph{copying}, $C$ and \emph{uncascading}, $B^{-1}$.  Finally, we reverse the sort and map the data back to the original registers. Accordingly, the parallel look-up algorithm can be written
\begin{eqnarray}
  U_{(N,N)} &=& \cheese{F} \circ S_{2N}^{-1} \circ B^{-1} \circ C \circ B \circ S_{2N} \circ F.
\end{eqnarray}

\paragraph{1) Initial Formatting}

For the parallel look-up algorithm, we use packets containing four items. Each packet contains an address and flag as before, but now we have two data registers of $d$ bits;  target-data $y_i$ and memory-data $x_i$. The initial formating stage, $F$, resembles the same step in Eq. (\ref{eq:format}) but where packet locations are initially stored as quantum data,
\begin{eqnarray}
  F~:~\lefteqn{ \ket{j_1,\ldots,j_{N}} \ket{y_1,\ldots,y_{N}} \ket{x_1,\ldots,x_{N}} \bigotimes_{i=1}^{N} \ket{(0,0,0,0)}_{2i}~\ket{(0,0,0,0)}_{2i+1} }   \\
  &\mapsto&  \ket{0,\ldots,0} \ket{0,\ldots,0} \ket{0,\ldots,0} \bigotimes_{i=1}^{N} \ket{(i,a,0,x_i)}_{2i}~\ket{(j_i,q,y_i,0)}_{2i+1},  \nonumber
\end{eqnarray}
The map $F$ moves the data into the packets, where it can be processed by the rest of the algorithm.  The initial formating step is achieved in one time-step.

\paragraph{2) Sorting}

The sorting step is the same as before: we sort lexicographically reading only the address and flag of each packet.  
\begin{eqnarray}
  \lefteqn{ S_{2N} ~:~ \ket0 \otimes \bigotimes_{k=1}^{2N} \ket{(i_k,f_k,y_k,x_k)}_{k} } \phantom{XXX} \\
    &\mapsto&  \ket{\sigma} \otimes \bigotimes_{k=1}^{2N} \ket{(i_{\sigma(k)},f_{\sigma(k)},y_{\sigma(k)},x_{\sigma(k)})}_{k}.  \nonumber
\end{eqnarray}

Note that packets whose flag is $\ket{a}$ hold data in their memory-data registers, while packets whose flag is $\ket{q}$ hold `target data' in their target-data registers. At the end of the sort, we are left with a sequence of the form
\begin{eqnarray}\label{sorteddata}
  \ldots(i,q,y_*,0)\ldots (i,q,y_*,0)(i,a,0,x_{i})\ldots
\end{eqnarray}
where $*$ denotes the various processor that have queried processor $i$.  

\paragraph{3) Cascade}

The goal of \emph{cascade} is to send a copy of the data $x_i$ into the empty data registers of packets on the left (cf. \eqref{sorteddata}).  This can be done by performing CNOTs first to packets at distance 1,
then distance 2, 4, 8, etc.  The CNOTs operate on the fourth site of
each packet and are controlled to only act if the source and target packet both
have the same value of $j$.
Since there is no way of knowing in advance how far the data will need to propagate, we need a method that works in all cases. For example, it could be the case that every processor requests data from a single processor, say $j_1=j_2=\ldots=j_{N}=1$.

This can be achieved by dividing the cascade up into $n$ smaller \emph{phases}, where $n=\lceil \log_2(2N) \rceil$.  Accordingly we write
\begin{eqnarray}
  B  &=&  B_{n-1} \circ B_{n-2} \circ \ldots \circ B_1 \circ B_0,
\end{eqnarray}
where each $B_k$ phase acts on pairs of packets at relative index separation $2^k$ from one another, more or less in parallel. 

%

To achieve this transformation, each packet will need a fresh ancilla \emph{aux-phase} register of $\lceil \log_2 n \rceil$ bits.  This will be used to store a number indicating the phase in which that packet acquires a copy of the data being cascaded to it.  This record makes it much easier to implement the whole cascade reversibly, and these aux-phases are to be retained until the cascade is later reversed.  Each packet will also need an \emph{aux-action} bit that persists only throughout a single phase (recycled from one phase to the next), being set if that packet receives data during that phase.

Each phase involves many pairs of packets.  Phase $B_k$ involves pairs with indices $l$ and $l+2^k$, for all $l$ for which $l \ge 0$ and $l+2^k < 2N$.  If $\lfloor 2^{-k}l \rfloor$ is even (\emph{resp.} odd), then the pair $(l, l+2^k)$ is said to be \emph{of even (\emph{resp.} odd) parity}.
The idea is that data can legitimately cascade from the packet at $l+2^k$ to the one at $l$ if they have the same address and if the rightmost one presently has data but the leftmost one doesn't.  It is enough to check that the leftmost one has $\ket{q}$ for its flag and $\ket0$ for its aux-phase, and that the rightmost one has $\ket{a}$ for its flag or something non-zero for its aux-phase.  If that overall condition is met, flip the aux-action bit of the leftmost packet, because it will be receiving data this phase.  It is necessary to set the aux-actions for all the even parity pairs first, then for all the odd parity pairs, because otherwise the same bits will be being read by different gates at the same time, and that violates the rules of the standard circuit model.  Note also that the simple act of computing the aux-action bit may itself require a little extra ancilla scratch space. 

When \emph{all} the aux-action bits have been correctly set for the present phase, cascade data leftwards, locally conditioned on those aux-action bits.  For example, during phase $B_k$ when examining packets $l$ and $l+2^k$, if $l$ has had its aux-action set, then we need to implement
\begin{eqnarray}
  \lefteqn{ \ket{(i,q,y',0)}_l ~\ket{0}_{phase(l)} ~\otimes~ \ket{(i,f,y,x)}_{l+2^k} ~\ket{p}_{phase(l+2^k)} }  \phantom{XXXX} \\ 
  &\mapsto&  \ket{(i,q,y',\red{x})}_l ~\ket{\red{k+1}}_{phase(l)} ~\otimes~ \ket{(i,f,y,x)}_{l+2^k} ~\ket{p}_{phase(l+2^k)}. \nonumber
\end{eqnarray}
(The aux-action for $l$ got set because one of $f$ and $p$ was non-zero.)
This should be done for the even parity pairs first (say), then the odd parity ones.

Finally, the aux-action bits need to be reset.   This resetting is a `local' operation, because during phase $B_k$, each packet need flip its aux-action if and only if its aux-phase is $\ket{k+1}$, indicating that it was active this phase.  
Note that the condition for resetting an aux-action bit is completely different from the condition for setting it in the first place.
 
The total effect of $B$ will be to load up the aux-phase ancillas and to replace every instance of $\ket{(j_i,q,y_i,0)}$ with $\ket{(j_i,q,y_i,x_{j_i})}$, while preserving every instance of $\ket{(i,a,0,x_i)}$.

\paragraph{4) Copying}

$C$ is a simple depth 1 local operation.  Every packet simply CNOTs the contents of its memory-data into its target-data.  
This has the effect of mapping every $\ket{(j_i,q,y_i,x_{j_i})}$ to $\ket{(j_i,q,y_i \red{\oplus x_{j_i}},x_{j_i})}$, while every $\ket{(i,a,0,x_i)}$ gets mapped to $\ket{(i,a,\red{x_i},x_i)}$.

\paragraph{5) Reversing the Cascade}

This map reverses the effect of the cascade, cleaning up all the aux-phase ancillas, making the packets ready for unsorting.

The total effect of $\left( B^{-1} \circ C \circ B \right)$ is to replace every instance of $\ket{(j_i,q,y_i,0)}$ with $\ket{(j_i,q,y_i \oplus x_{j_i},0)}$, while replacing $\ket{(i,a,0,x_i)}$ with $\ket{(i,a,x_i,x_i)}$ for all $i$.  This action does not change the ordering of the packets, which only depends on the address and flag of each packet.  Therefore the action is compatible with the sorting stages, as required.

\paragraph{6) Unsort}

The unitary $S_{2N}^{-1}$ unsorts just as before.

\paragraph{7) Final Formatting}

The final step is to apply a formatting map, $\cheese{F}$, which works as follows,
\begin{eqnarray}
  \lefteqn{ \ket{0,\ldots,0} \ket{0,\ldots,0} \ket{0,\ldots,0} \bigotimes_{i=1}^{N} \ket{(i,a,x_i,x_i)}_{2i}~\ket{(j_i,q, y_i \oplus x_{j_i},0)}_{2i+1}}   \\
  &\mapsto& \ket{j_1,\ldots,j_{N}} \ket{y_1\oplus x_{j_1},\ldots,y_{N}\oplus x_{j_N}} \ket{x_1,\ldots,x_{N}} \bigotimes_{i=1}^{N} \ket{(0,0,0,0)}_{2i}~\ket{(0,0,0,0)}_{2i+1}.  \nonumber
\end{eqnarray}
Note that this is a depth 2 map rather than a depth 1 map because each $x_i$ appears in two places, and these need to `relocalise' as well as `move'.


\section{Revisiting Popular Algorithms}  \label{sec:algs}

Grover's quantum search algorithm \cite{Grover96} and the generalization to amplitude amplification \cite{Brassard+97, Brassard+00} have the great advantage of being widely applicable. Problems such as Element Distinctness \cite{buhrman+00, ambainis07}, Collision Finding \cite{Brassard+97b}, Triangle Finding \cite{belovs12, magniez+07}, 3-SAT \cite{ambainis04} and NP-hard tree search problems \cite{cerf+00} all have faster quantum algorithms because they are able to make use of amplitude amplification. In this section, we revisit Grover's quantum search algorithm and the Element Distinctness problem, resolving a challenge posed by Grover and Rudolph \cite{grover+rudolph}.

The theorems presented in this section use the the circuit model so that they
can be easily compared to the existing literature. However, we developed the ideas using the quantum parallel RAM model and indeed, we borrow some of the terminology in the theorem proofs.

The results in this section follow from applying our main result (\thmref{main}) to existing quantum algorithms by Grover \cite{Grover96} and Buhrman et al. \cite{buhrman+00}.

\subsection{A single quantum search of an unstructured database}
\label{sec:Grover}

Grover's fast quantum search algorithm \cite{Grover96} is usually presented as solving an oracle problem: let $f:\{1,\ldots ,N\} \rightarrow \{0,1\}$ be a function and suppose we wish to find solutions, $s$, such that $f(s)=1$. We are given an oracle with the ability to recognise solutions in the form of a unitary 
\begin{eqnarray}
  U_f  &:&  \ket{x}\ket{y} ~~\mapsto~~ \ket{x}\ket{y\oplus f(x)}.
\end{eqnarray}
%
Setting the target register of $U_f$ to the state $(\ket0-\ket1)/\sqrt{2}$ encodes the value of $f(x)$ into a phase shift
\begin{eqnarray}
  U_f  &:&  \ket{x} ~~\mapsto~~ (-1)^{f(x)} \ket{x},
\end{eqnarray}
where we have suppressed the target register since it remains unchanged.
Grover's algorithm then makes $\Theta(\sqrt{N/M})$ calls to $U_{f}$ and with probability $1-O(M/N)$ outputs one of the $M$ possible solutions uniformly at random. 

Grover's algorithm can be applied to search an unstructured database. We construct the oracle by using the single memory look-up unitary, $U_{(1,N)}$, together with a simple function that tests whether a database entry is a solution (see Chap.~6.5 in \cite{nielsen+chuang}). More generally, suppose we wish to find solutions to a function whose inputs are not expressed simply as the numbers from $1$ to $N$, but rather are taken from a database of elements $X = \{ x_{j}:j=1,\ldots,N \},$ where each $x_{j}$ is a bit string of length $d$. That is, we have a function $\alpha:X \rightarrow \{0,1\},$ and we are searching for solutions $s \in \{1,\ldots,N\}$ such that $\alpha(x_{s})=1$. Once the database item $x_j$ has been loaded into the computational memory, the function $\alpha$ is computed using the unitary  
\begin{eqnarray}
  U_\alpha  &:&  \ket{x_j}\ket{y} ~~\mapsto~~ \ket{x_j}\ket{y\oplus \alpha(x_j)}.
\end{eqnarray}

We consider the case in which the database is held in a quantum state $\ket{x_1, \ldots, x_N }$, but it could also be a classical database whose indices can be accessed in superposition \cite{Giovannetti+08, Giovannetti+08b}.

An oracle is constructed by first looking-up a database entry and then testing if this entry is a solution to the function $\alpha$. The initial state of the computer is
\begin{eqnarray}
\ket{j}\ket{0}\ket{y}\ket{x_1, \ldots, x_N }
\end{eqnarray}
where the state $\ket{j}$ is the index of the database item we will look-up, $\ket{0}$ is used to load the memory and $\ket{y}$ is the target register.

First we apply the single memory look-up unitary, $U_{(1,N)}$, (see Definition~\ref{defn:database}) so that the computer is in the state
\begin{eqnarray}
\ket{j}\ket{x_j}\ket{y}\ket{x_1, \ldots, x_N },
\end{eqnarray}
then calling the function $\alpha$, using the corresponding unitary $U_\alpha$, maps the state to
\begin{eqnarray}
\ket{j}\ket{x_j}\ket{y\oplus \alpha(x_j)}\ket{x_1, \ldots, x_N } .
\end{eqnarray}
Finally we restore the auxiliary state used to load the database item by applying $U_{(1,N)}^\dagger=U_{(1,N)}$. The final state of the computation is therefore
\begin{eqnarray}
\ket{j}\ket{0}\ket{y\oplus \alpha(x_j)}\ket{x_1, \ldots, x_N } .
\end{eqnarray}
Hence the unitary 
\begin{eqnarray}
  \mathcal{O}_\alpha = U_{(1,N)} \circ U_\alpha \circ U_{(1,N)}
\end{eqnarray}
can be used as an oracle for Grover's algorithm. Setting the target register $\ket{y}=(\ket0-\ket1)/\sqrt{2}$ encodes the value of $\alpha(x_j)$ into a phase shift
\begin{eqnarray}
  \mathcal{O}_\alpha  &:&  \ket{j} ~~\mapsto~~ (-1)^{\alpha(x_j)} \ket{j}.
\end{eqnarray}
From \thmref{lower bound}, the quantum circuit implementing Grover's algorithm with $\mathcal{O}_\alpha$ as an oracle requires circuit
width $\Omega(Nd)$
and depth $\Omega(\sqrt{N/M}(\log N+D_{\alpha})$ to find one solution with high probability.  Here $D_{\alpha}$ is the depth of the circuit for $U_{\alpha}$.

In many cases of interest (cf. \secref{EltDist}), we actually want to search over pairs (or more generally $r$-tuples) of database elements.  We thus generalise the preceding algorithm by considering a function $\alpha:X^r\to\{0,1\}$ that takes as input an $r$-tuple of database entries.  We wish to find solutions $\mathbf{s}=(s^1,\ldots,s^r)\in\{1,\ldots,N\}^r$ such that $\alpha(x_{s^1},\ldots,x_{s^r})=1$.  As with the case $r=1$, we construct an oracle $\mathcal{O}_{\alpha}$ for Grover's algorithm using $U_{(1,N)}$ and the unitary
\begin{eqnarray}
    U_\alpha  &:&  \ket{\mathbf{x_j}}\ket{b} ~~\mapsto~~ \ket{\mathbf{x_j}}\ket{b\oplus \alpha(\mathbf{x_j})},
\end{eqnarray}
where $\mathbf{j}=(j^1,\ldots,j^r)\in\{1,\ldots,N\}^r$ and $\mathbf{x_j}=(x_{j^1},\ldots,x_{j^r})\in X^r$.  The construction of the oracle runs just as in the previous case, but with the first register labelled with the multi-index $\mathbf{j}$, while the second register contains $r$-tuples of database entries $\mathbf{x_j}$.  The initial state is
\begin{eqnarray}
  \ket{\mathbf{j}}\ket{\mathbf{0}}\ket{b}\ket{x_1, \ldots, x_N }.
\end{eqnarray}
Applying the single memory-lookup $U_{(1,N)}$ $r$ times, once to each label of the multi-index and each position in the memory-register, we obtain the state
\begin{eqnarray}
  \ket{\mathbf{j}}\ket{\mathbf{x_j}}\ket{b}\ket{x_1, \ldots, x_N }.
\end{eqnarray}
Next we call the unitary $U_{\alpha}$ to arrive at the state
\begin{eqnarray}
  \ket{\mathbf{j}}\ket{\mathbf{x_j}}\ket{b\oplus\alpha(\mathbf{x_j})}\ket{x_1, \ldots, x_N }
\end{eqnarray}
and finally we uncompute the $r$ memory-lookup steps to obtain
\begin{eqnarray}
  \ket{\mathbf{j}}\ket{\mathbf{0}}\ket{b\oplus\alpha(\mathbf{x_j})}\ket{x_1, \ldots, x_N }.
\end{eqnarray}
Setting the target register to $(\ket0-\ket1)/\sqrt{2}$ we obtain the oracle
\begin{eqnarray}
    \mathcal{O}_\alpha  &:&  \ket{\mathbf{j}} ~~\mapsto~~ (-1)^{\alpha(\mathbf{x_j)}} \ket{\mathbf{j}}.
\end{eqnarray}
The quantum circuit implementing Grover's algorithm with $\mathcal{O}_{\alpha}$ as an oracle makes $\Theta(\sqrt{N^r/M})$ calls to $\mathcal{O}_{\alpha}$, where $M$ is the number of solutions.  From \thmref{lower bound}, the circuit width and depth are $\Omega(Nd)$ and $\Omega(\sqrt{N^r/M}(r\log N+D_{\alpha}))$, respectively.

\subsection{Parallel quantum search of an unstructured database}
Now that we have an efficient algorithm for performing parallel memory look-ups, we consider the effect of using the unitary $U_{(N,N)}$ together with (up to) $N$ functions as oracles for Grover's algorithm. We show how multiple quantum processors can each run a Grover search on a single database \emph{in parallel}. Since the first step of each instance of Grover's algorithm is to query the entire database, it may come as a surprise that we do not encounter (unwanted) interference. It is the efficiency of our parallel memory look-up algorithm that allows us to interleave the Grover steps with the parallel database access.

Suppose we have (up to) $N$ functions that take $r$ database elements as inputs, $\alpha_{i}:X^r \rightarrow \{0,1\},$ for $i=1,\ldots N$.  We wish to find solutions $\mathbf{s_i}=(s_i^1,\ldots,s_i^r)\in\{1,\ldots,N\}^r$ such that $\alpha_{i}(x_{s_i^1},\ldots,x_{s_i^r})=1$ for all $i=1,\ldots,N$.  In exact analogy to the previous section, we assume that we are given circuits for the unitaries
\begin{eqnarray}
   U_{\alpha_i}  &:&  \ket{\mathbf{x_j}}\ket{b} ~~\mapsto~~ \ket{\mathbf{x_j}}\ket{b\oplus \alpha_i(\mathbf{x_j})},
\end{eqnarray}
where again we write $\mathbf{j}=(j^1,\ldots,j^r)\in\{1,\ldots,N\}^r$ and $\mathbf{x_j}=(x_{j^1},\ldots,x_{j^r})\in X^r$.  We shall assume that the depth and width of the circuits for $U_{\alpha_i}$ are polynomial functions of $\log N,d$ and $r$.  In effect we are interested in functions that are easy to compute but we do not know how to invert. Thus, the best method for finding a solution is to treat the function as a black box. Each node has size $d$, which is $\Omega(\log N)$ since it must be large enough to perform basic operations, hold the database items and perform function evaluations. Thus we assume that $d$ is a polynomial in $\log N$.

With these assumptions, the following theorem provides an algorithm that finds a solution for \emph{each} function $\alpha_i$, using the same size quantum circuit (up to a factor polynomial in $\log N$ and $r$) required to find only one solution using the unitary $U_{(1,N)}$ as an oracle (see \secref{Grover}).

The algorithm can be expressed entirely in terms of the circuit model but we find it convenient to talk of quantum processors and memory locations. Now that we know the quantum parallel RAM model can be emulated by the circuit model with a $O(\log N)$ overhead, the algorithm presented below becomes a simple generalisation of Grover's algorithm. The number of processors, $N$, equals the number of items in the database. Reiterating the arguments of Grover and Rudolph \cite{grover+rudolph}, we anticipate that quantum memory will cost the same as quantum processors.



\begin{theorem}[Multi-Grover search algorithm] \label{thm:multiGrover}

Using the notation defined above, there is a quantum algorithm that for each $i=1,\ldots,N$ either returns the solution index $\mathbf{s_{i}}\in\{1,\ldots,N\}^r$ such that $\alpha_{i}(\mathbf{x_{s_{i}}})=1$, or, if there is no such solution, returns `no solution'.  The algorithm succeeds with probability $\Theta(1)$ and can be implemented using a quantum circuit with width $\tilde{O}(N)$ and depth $\tilde{O}(\sqrt{N^r/M})$.
\end{theorem}
The value $M$ in the statement of \thmref{multiGrover} is the minimum over $i\in\{1,\ldots,N\}$ of the number of solutions $\mathbf{s_i}$ to $\alpha_i(\mathbf{x_{s_i}})=1$.  The notation $\tilde{O}$ means that we are neglecting a multiplicative factor that is polynomial in $r\log N$.

\emph{Proof~:~}
First, the case $r=1$.  We proceed as with the single Grover search of an unstructured database, the only subtlety being that we need to organise the Hilbert space in the correct way. We want to perform $N$ Grover searches over the database so we need $N$ indices $\ket{j_1 \ldots j_N}$, $N$ memory place holders $\ket{0 \ldots 0}$ and $N$ target registers $\ket{b_1 \ldots b_N}$.

It is useful to split the circuit in to $N$ `processors' since we will think of each one as performing a Grover search over the database. We rearrange the Hilbert space across $N$ processors as
\begin{eqnarray}  
  \ket{j_1,\ldots, j_{N}} \ket{0,\ldots, 0} \ket{b_1,\ldots, b_{N}} \ket{x_1,\ldots, x_{N}} 
    &\cong&  \bigotimes\limits_{i=1}^{N} \ket{j_i,0,b_i,x_i}.
\end{eqnarray}
Applying the parallel look up algorithm $U_{(N,N)}$, maps the initial state of the computer to 
\begin{eqnarray}  
  \bigotimes\limits_{i=1}^{N} \ket{j_i,x_{j_i},b_i,x_i}.
\end{eqnarray}

We define a circuit for implementing the unitaries $U_{\alpha_{i}}$ in parallel as
\begin{eqnarray}
  U_{\alpha}  &=&  \bigotimes\limits_{i=1}^{N} U_{\alpha_{i}},
\end{eqnarray}
which acts on all of the target register simultaneously sending the state to 
\begin{eqnarray}  
  \bigotimes\limits_{i=1}^{N} \ket{j_i,x_{j_i},b_i\oplus \alpha_i(x_{j_i}),x_i}.
\end{eqnarray}

Finally, we clear the auxiliary register used to load the memory item using the unitary $U_{(N,N)}$, so that the final state of the computer is 
\begin{eqnarray}  
  \bigotimes\limits_{i=1}^{N} \ket{j_i,0,b_i\oplus \alpha_i(x_{j_i}),x_i}.
\end{eqnarray}
Setting each of the target registers to $(\ket0-\ket1)/\sqrt{2}$ produces the required oracle using a circuit with width $O(N(d+\log{N}+W_{\alpha}))$ and depth $O(\log N\log(d\log N)+D_{\alpha})$  (c.f. \thmref{main}).  Here $W_{\alpha}$ and $D_{\alpha}$ are the maximum over $i\in\{1,\ldots,N\}$ of the widths and depths of the circuits $U_{\alpha_i}$, respectively.

Grover's algorithm then calls the unitary
\begin{eqnarray}  
  U_{(N,N)} \circ U_\alpha \circ U_{(N,N)}
\end{eqnarray}
$O(\sqrt{N/M})$ times, where $M$ is the minimum over $i\in\{1\,\ldots,N\}$ of the number of solutions of $\alpha_i$. Note that if the number of calls varies for different functions $\alpha_i,$ we simply pad-out the oracle calls for the respective $\alpha_i$  using the identity. The resulting algorithm finds a solution for each function $\alpha_i$ using a circuit of width $O(NW_{\alpha})=\tilde{O}(N)$ and depth $O(\sqrt{N/M}(\log{N}\log d+D_{\alpha}))=\tilde{O}(\sqrt{N/M})$.

The case $r>1$ is a straight-forward generalisation, replacing the indices $j_i$ and memory-data $x_{j_i}$ by $r$-tuples $\mathbf{j_i}$ and $\mathbf{x_{j_i}}$, respectively, and replacing each single call of $U_{(N,N)}$ by $r$ calls, in direct analogy to the previous section.  We obtain a unitary $\mathcal{O}_{\alpha}$ with width $O(N(rd+r\log{N}+W_{\alpha}))$ and depth $O(r\log N\log(d\log N)+D_{\alpha})$.

Grover's algorithm makes $O(\sqrt{N^r/M})$ calls to $\mathcal{O}_{\alpha}$ and the resulting circuit has width 
$$O(N(rd+r\log{N}+W_{\alpha}))=\tilde{O}(N)$$
and depth 
$$O(\sqrt{N^r/M}(r\log N\log(d\log N)+D_{\alpha})=\tilde{O}(\sqrt{N^r/M}) \, ,$$
as required.
\hfill  \qed  \bigskip

If $X$ were highly structured (such as being the numbers 1 to $N$), it would be straightforward to perform $N$ Grover searches in parallel, using a circuit of width $\tilde{O}(N)$, since we would not need to store $X$ explicitly.  Now, making use of the efficient parallel memory look-up algorithm to access $X$, we are able to interlace the steps in the Grover algorithm with database look-ups. The end result is that we can indeed perform $N$ Grover searches in parallel regardless of the structuring of $X$. We now examine the effect of \thmref{main} on other memory intensive quantum algorithms.

\subsection{Element Distinctness}  \label{sec:EltDist}

In this section, we present a quantum algorithm for the Element Distinctness problem: given a function $f:\{0,1\}^n \rightarrow \{0,1\}^{n}$, determine weather there exist distinct $i,j\in \{0,1\}^n$ with $f(i)=f(j)$. It is based on the algorithm of Buhrman et al. \cite{buhrman+00} but improved by making use of our efficient algorithm for parallel memory look-ups (\thmref{main}).

The size of the problem is parametrised by $N=2^{n}$.  Let $S$ and $T$ denote the available memory and time respectively. Classically we can search for a solution to the Element Distinctness problem using $O(N)$ function calls provided $ST=O(N^2)$.  Applying Grover's algorithm to this classical search fails to achieve an improvement in the number of function queries: there are $\Theta(N^2)$ pairs to search, so Grover's algorithm requires $\Theta(N)$ queries.
Ambainis~\cite{ambainis07} discovered an elegant quantum algorithm that improves upon this, having query complexity $O(N^{2/3})$. In fact $\Omega (N^{2/3})$ queries is also a lower bound, so Ambainis's algorithm is known to be optimal from this perspective. However, Ambainis requires memory proportional to $\tilde{O}(N^{2/3})$; i.e. $S=T=\tilde{O}(N^{2/3})$.

If we think of $f$ as something that is computed by a small (say
$\poly\log N$ size) circuit, then the same performance as Ambainis's
algorithm can be achieved by a much more naive
approach~\cite{grover+rudolph}.  We simply divide the $N^2$ search
space into $N^2/S$ blocks and Grover search each one in time
$\tilde{O}(\sqrt{N^2/S})$ for a time-space trade-off of $ST^2=\tilde{O}(N^2)$, which includes
both Ambainis's special case $S=T=\tilde{O}(N^{2/3})$ and the naive Grover search $S=O(\poly \log N),T=\tilde{O}(N)$.

We now show how to improve on these results in the DQC model answering a challenge posed by Grover and Rudolph \cite{grover+rudolph}. We are interested in the case where $f$ can be cheaply evaluated, but not reverse-engineered (so that our best method of finding a collision is to treat $f$ as a black box). 

\begin{theorem}  \label{thm:EDtrade-off}
Given a $\poly \log N$ size circuit for computing the function $f$, we can solve the Element Distinctness problem in space $\tilde{O}(S)$ and time $T=\tilde{O}(N/S)$ for any $S \le N$. This results in the space-time trade-off 
$$
ST=\tilde{O}(N) \, .
$$
\end{theorem}

\emph{Proof~:~}
First choose a random set of $S$ inputs $x_1,\ldots,x_S$ to query and 
store these along with $L=\{ f(x_1),\ldots,f(x_S) \}$ in the distributed memory. That is, processor $j$ stores the pair $(x_j, f(x_j))$. 
Next, we use the $S$ processors to check the 
remaining $N-S$ inputs for a collision with one of the
$x_1,\ldots,x_S$. 

Using the AKS sorting network, the $S$ processors can sort $L$ in time $O(\log N)$. It is now easy to check if $L$ contains any pair of elements that are equal, $f(x_i)=f(x_j)$. If we find such a pair then output $(i,j)$.  Otherwise use the sorted list, $L$, to construct a function, $g_L:\{1, \ldots, N-S\}\rightarrow \{ 0,1 \}$, defined by 
\begin{eqnarray}
g_L(i)=\left\{ 
\begin{array}{cl}
1 & \mbox{ if there exists an element $j \in \{1,\ldots,S\}$ such that $f(x_i)=f(x_j)$}\\ 
0 & \mbox{ otherwise.}%
\end{array}%
\right. 
\end{eqnarray}

Now we can split the remaining $N-S$ inputs into $S$ collections of size $O(N/S)$ and use the Multi-Grover algorithm to perform $S$ parallel searches, each with the function $g_L$ as an oracle. 
This procedure requires $\Theta(\sqrt{N/S})$ calls to $g_L$. The cost of calling the function $g_L$ in parallel is $O(\log S \log N \log \log N)$ using a binary search combined with our algorithm $U_{(S,S)}$ for making simultaneous queries to a distributed memory.
Hence the total depth is $\tilde{O}(\sqrt{N/S})$.

We will succeed (with constant probability) in finding a collision
if one of the colliding elements was in $\{x_1,\ldots,x_S\}$.  Since
this event happens with probability $\Theta(S/N)$, we can wrap the
entire routine in amplitude amplification to boost the probability to
$\Theta(1)$ by repeating $\Theta(\sqrt{N/S})$ times.  Hence the total runtime is 
$T=\tilde{O}\left(N/S  \right)$ and so we have achieved a time-space trade-off of 
$ST = \tilde{O}(N) $.
\hfill  \qed  \bigskip

\subsection{Collision Finding problem}

Our results also apply to the Collision Finding problem: in which $f:[N]\mapsto [N]$ is an efficiently-computable function with the promise of being either 1-1 or 2-1, for which an $ST=O(N^{2/3})$ query algorithm is given in \cite{Brassard+97b}.  This problem may be solved with
$$ST=\tilde{O}(\sqrt{N})$$
either by selecting $O(\sqrt{N})$ random elements and solving Element Distinctness, or by simply using the algorithm of \cite{Brassard+97b} directly, augmented by using $S$ processors with shared memory together with our look-up algorithm\footnote{Bernstein~\cite{bernstein09} discusses time/space tradeoffs for a related problem in other computational models, including a 2-d nearest-neighbour architecture and a variant of the quantum RAM model that he refers to as ``a naive model of communication.''  Ref.~\cite{bernstein09} also uses a distributed sorting algorithm to parallelise collision finding, which could be seen as a special case of our main result.}.
So we perform $S$ Grover searches in parallel on spaces of size $N/S^2$ instead of a single search on a space of size $N/S$.

%

\section{Conclusion and discussion}  \label{sec:conclusion}
In classical parallel computing, sorting networks provide an elegant solution to the routing problem and simulation of the parallel RAM model. In this paper, we have demonstrated that they can be applied to quantum computing too.

The information about the connectivity of a quantum circuit is available before we run the algorithm (at compile time). Using this classical information we have designed an efficient scheme for routing quantum packets. The application of this data-moving algorithm is to distributed quantum computing. We provide an efficient way of mapping arbitrary unconstrained circuits to limited circuits respecting the locality of a graph. 

Our results already apply to nearest neighbour architectures in the case of a circuit that is highly parallel. The case of emulating a circuit with many concurrent operations on a 1D nearest neighbour machine was covered by Hirata et al. \cite{hirata}. The approach is to use the Insertion/Bubble sort to perform all of the operations in $O(N)$ time-steps which compares favorably to performing each gate in turn in $O(N^2)$ depth. We put this idea in a general framework applying to any (connected) graph. Along the way we are able to prove that up to polylogarithmic factors, this approach is optimal.

We have shown how the addition of  a few long-range (or flying) qubits dramatically increases the power of a distributed quantum computer. Using only $O(\log N)$ connections per node enables efficient sorting over the hypercube. A distributed quantum computer with nodes connected according to the hypercube graph would be able to emulate arbitrary quantum circuits with only $O(\log^2 N)$ overhead. One might expect that a quantum computer requires $O(N)$ connections per node so that each qubit can potentially interact with any other qubit. Our result demonstrates that this is not the case: for a small overhead $O(\log N)$ connections suffice.

We have presented a new algorithm for accessing quantum memory in parallel. The algorithm is a modification of the data-moving algorithm used in Sections~\ref{sec:processors} and \ref{sec:DQC} but where the destinations are \emph{quantum} data and no longer restricted to form a permutation. The algorithm is extremely efficient; it has an overhead that is scarcely larger than \emph{any} algorithm capable of accessing \emph{even a single entry} from memory. \thmref{main} implies that $N$ processors can have unrestricted access to a shared quantum memory. It tells us that the quantum parallel RAM and the circuit models are equivalent up to logarithmic factors.

Finally, we demonstrated that the parallel look-up algorithm can be used to optimize existing quantum algorithms.  We provided an extension of Grover's algorithm that efficiently performs multiple simultaneous searches over a physical database, and answered an open problem posed by Grover and Rudolph by demonstrating an improved space-time trade-off for the Element Distinctness problem. 
It seems likely that this framework for efficient communication in parallel quantum computing will be a useful subroutine in other memory-intensive quantum algorithms, such as triangle finding, or more generally for frameworks such as learning graphs.

\section*{Acknowledgments}
The authors would like to thank the Heilbronn Institute for Mathematical Research for hosting the discussions that led to this research.  We are grateful to Frederic Magniez and Tony Short for catching bugs in a previous version and an anonymous referee for highlighting relevant references.

AWH was funded by NSF grants 0916400, 0829937, 0803478, DARPA QuEST
contract FA9550-09-1-0044 and IARPA via DoI NBC contract D11PC20167. The U.S. Government is authorized to reproduce and distribute reprints for Governmental purposes notwithstanding any copyright annotation thereon. The views and conclusions contained herein are those of the authors and should not be interpreted as necessarily representing the official policies or endorsements, either expressed or implied, of IARPA, DoI/NBC, or the U.S. Government.

\section*{Proof of \thmref{main}}  \label{sec:count}

The cost of our parallel look-up algorithm presented in \thmref{main} and \secref{descU} is specified in terms of the circuit model. 
A circuit is decomposed (perhaps recursively) into a series of subroutines.  
In the circuit model, every subroutine is built out of gates from a universal gate set 
(cf. for example, reference \cite{nielsen+chuang}). 
Gates can be implemented concurrently---that is, within the same \emph{timeslice}---whenever they act on disjoint sets of qubits. The \emph{cost} of a circuit is measured by three parameters:
\begin{itemize}
  \item  \emph{depth}, the number of timeslices required in the circuit;
  \item  \emph{size}, the total number of gates in the circuit, and
  \item  \emph{width}, the total number of qubits (inputs + ancillas) required in the circuit.
\end{itemize}
These three are all taken to be functions of the \emph{logical unitary width}, which is the number of input qubits required for the logical unitary. 

%

In \secref{descU}, we provide circuits that implement the logical unitary $U_{(N,N)}$.  Here we count up the total depth and width of the circuit thereby proving \thmref{main}. 

The total depth of the formatting subroutines is $O(1)$.

The sorting subroutines require comparators that make a lexicographic comparison on $\lceil \log_2 N \rceil + 1$ bits, and lexicographic comparison is basically the first part of \emph{arithmetic subtraction} (regard the bit-patterns as integers, subtract one from the other reversibly, read the sign bit of the output).  This can be achieved efficiently in depth $O(\log \log N)$ \cite{Thapliyal+10}.  The other thing a comparator does is to swap elements controlled on a sorting-bit.  The sizes of our elements are $O( \log N + d )$, and so the sorting bit needs to be fanned out (using a binary tree) to `copy' it across $O( \log N + d )$ ancillas, before a depth 1 swap can take place.  Therefore the swapping stage of a comparator takes depth $O( \log( \log N + d ) )$, and so the total comparator has depth $O( \log \log N + \log( \log N + d ) ) = O( \log( d \log N ) )$.  Since there are $O( \log N )$ comparators in the AKS sorting network, the total depth of the sorting stage is $O( \log N \cdot \log( d \log N ) )$.

The cascade subroutine has $n = O(\log N)$ phases, and as with the comparators used in sorting, each phase of cascade involves arithmetic (in fact equality testing) on objects of size $O(n)$, plus controlled copying of objects of size $O(d)$.  Therefore a phase has depth $O( \log n + \log d) = O( \log( d \log N ) )$ and the total depth of the cascade part is $O( \log N \cdot \log( d \log N ) )$.

The inner subroutine (`copying') of the algorithm for $U_{(N,N)}$ has depth $O(1)$ and requires no ancillas. Therefore the total depth of our algorithm for $U_{(N,N)}$ is $O( \log N \cdot \log( d \log N ) )$.

Counting bits, we see that besides the $O( N (\log N+d) )$ input/output bits, we require $O( N (\log N+d) )$ more bits for the packets, $O(N \log N)$ sorting-ancilla bits, as well as $O( N (\log N+d) )$ ancilla bits for temporary use while rendering the (lexicographic) comparators.  
Furthermore, for \emph{cascade} each packet needs an aux-phase register and an aux-action bit (total $O(N \log N )$ bits), as well as $O((\log N +d) )$ scratch space (\emph{e.g.} to compute/reset the aux-action bit) that can be recycled between phases. 
Hence, the total circuit width is $O(N (\log N+d) )$, which is linear in the width of the logical unitary in question, and therefore asymptotically optimal.
\hfill  \qed

\end{document}